\documentclass[11pt,oneside,english]{amsart}
\usepackage[T1]{fontenc}
\usepackage[latin9]{inputenc}
\usepackage[letterpaper]{geometry}
\usepackage{textcomp}
\usepackage{amsthm}
\usepackage{amssymb}
\usepackage{esint}

\makeatletter
\numberwithin{equation}{section}
\numberwithin{figure}{section}

\def\makebbb#1{
    \expandafter\gdef\csname#1\endcsname{
        \ensuremath{\Bbb{#1}}}
}\makebbb{R}\makebbb{N}\makebbb{Z}\makebbb{C}\makebbb{H}\makebbb{E}\makebbb{D}\makebbb{P}\makebbb{B}\makebbb{K}\makebbb{E}

\usepackage{babel}

\usepackage{babel}

\makeatother

\usepackage{babel}

\makeatother

\usepackage{babel}

\begin{document}

\title{Kähler-Einstein metrics emerging from free fermions and statistical
mechanics }

\author{Robert J. Berman}

\address{Chalmers University of Technology and University of Gothenburg}

\email{robertb@chalmers.se}
\begin{abstract}
We propose a statistical mechanical derivation of Kähler-Einstein
metrics, i.e. solutions to Einstein's vacuum field equations in Euclidean
signature (with a cosmological constant) on a compact Kähler manifold
$X.$ The microscopic theory is given by a canonical free fermion
gas on $X$ whose one-particle states are pluricanonical holomorphic
sections on $X$ (coinciding with higher spin states in the case of
a Riemann surface) defined in background free manner. A heuristic,
but hopefully physically illuminating, argument for the convergence
in the thermodynamical (large $N)$ limit is given, based on a recent
mathematically rigorous result about exponentially small fluctuations
of Slater determinants. Relations to higher-dimensional effective
bosonization, the Yau-Tian-Donaldson program in Kähler geometry and
quantum gravity are explored. The precise mathematical details will
be investigated elsewhere.
\end{abstract}
\maketitle

\section{Introduction}

The basic laws of gravity have an intriguing similarity with the laws
of thermodynamics and hydrodynamics - this has been pointed out at
several occasions in the physics literature, in particular in connection
to the study of black holes (see for example \cite{bch,ja,v}). As
a consequence one is lead to ask whether gravity can be seen as an
emergent effect of an underlying microscopic theory in a thermodynamical
limit \cite{ja}? The aim of this note is to propose a situation where
this question can be answered in affirmatively. We will consider Einstein's
vacuum field equations in Euclidean signature on a compact manifold
$X$, whose solutions are usually called \emph{Einstein metrics} in
the mathematics literature \cite{an}. More precisely, these equations
will be considered in the presence of a fixed background integrable
complex structure $J$ on $X.$ It turns out that the underlying microscopic
theory may then be realized as a certain free fermion gas on $X$
(whose definition only involves $J$ but no metric data) and it will
be shown how to recover an Einstein metric (with a non-zero cosmological
constant) in the thermodynamical limit. The metric is singled out
by the fact that it is Hermitian and Kähler with respect to $J$.
In other words these are the \emph{Kähler-Einstein metrics} which
have been extensively studied during the last decade in the mathematics
literature (for a recent survey see \cite{p-s}). 

Physically, metrics as above appear, for example, as gravitational
instantons in Hawking's functional integral approach to quantum gravity
\cite{ha,ts}. Although we will not restrict $X$ to be a real four-manifold
- the physically most relevant case - it is worth pointing out that
in this latter case {}``most'' Einstein metrics are Kähler-Einstein
metrics. In particular, for a negative cosmological constant $\Lambda$
it may actually be that \emph{all} Einstein metrics are Kähler with
respect to\emph{ some} complex structure $J,$ as long as $X$ complex
structures (the question was raised in \cite{le2}). This is for example
the case for compact quotients of the unit ball, as shown in \cite{le}
using Seiberg-Witten gauge theory. By making the complex structure
$J$ dynamical, inspired by the works of Fujiki \cite{fu} and Donaldson
\cite{do3} in Kähler geometry, we will also explain how the microscopic
theory referred to above can be used to give a finite $N$ well-defined
approximation to a variant of quantum gravity (related to Liouville
gravity when $n=1$) where the role of the diffeomorphism group in
ordinary gravity is played by a symplectomorphism group which arises
as the gauge group for a bundle over the moduli space of all complex
structures on the smooth manifold $X.$ 

To be a bit more precise, given a complex structure $J$ on $X$ we
will obtain a canonical (metric free) probability  measure $\mu^{(N)},$
expressed in terms of a Slater determinant, on the $N-$ product $X^{N}$
such that in the large $N-$limit the measure $\mu^{(N)}$ becomes
exponentially concentrated on configurations of points approximating
the normalized volume form $\mu_{KE}$ of the Kähler-Einstein metric
of $(X,J),$ i.e. such that \[
\frac{1}{N}\sum_{i=1}^{N}\delta_{x_{i}}\approx\mu_{KE}\]
Since the corresponding Kähler-Einstein metric on $(X,J)$ is uniquely
determined by its volume form $\mu_{KE}$ this means that the Kähler-Einstein
metric indeed emerges macroscopically. Finally, by working over the
moduli space of all complex structures on $X$ the dependence on $J$
is also taken into account.

The main ingredients in the investigation of the thermodynamical limit
is the asymptotics of exponentially small fluctuations of Slater determinants
for $N-$particle correlations of fermions on complex manifolds in
\cite{berm3} (building on \cite{b-b,b-b-w,bbgz}). On one hand, from
a purely mathematical point of view these large $N$ asymptotics concern
large deviations for certain critical determinantal random point processes,
which generalize Random Matrix ensembles previously extensively studied.
On the other hand, from a physical point of view the result can be
seen as an effective bosonization of a free fermion gas (see section
\ref{sub:Derivation-of-}), which in the case of a Riemann surface
alternatively can be deduced from the exact bosonization results in
\cite{vv,b-v-}. The large deviation result for the Slater determinant
is then combined with a basic large deviation result for a non-interacting
\emph{classical} gas going back to Boltzmann's fundamental work on
entropy (called Sanov's theorem in the mathematics literature). 

It should however be pointed out that the argument in the present
note which combines the two mathematically rigorous results referred
to above is not completely rigorous. Basically, it involves an interchange
of two limits which needs to be mathematically justified. The mathematical
details, as well as various extensions, will be investigated elsewhere,
but hopefully the heuristic derivation given here is illuminating
from a physical point of view as it involves manipulations that are
standard in the functional integral approach to quantum field theory. 

Incidentally, in the case of a Riemann surface (i.e. the case when
the\emph{ real} dimension $D$ of $X$ is two) the situation studied
in the present note is closely related to the previous mathematical
study of various $2D$ ensembles (point vortex systems, plasmas, self-gravitating
systems, ...) from the point of view of mean field theory; see \cite{clmp,k}
and references therein. In particular, the corresponding thermodynamical
limit was studied in \cite{clmp,k} as a model of 2D turbulence. However,
the higher dimensional situation studied in the present work is analytically
considerably more complicated as the resulting limiting mean field
equations are fully non-linear (see section \ref{sub:The-minimizer-}).
The reason is that the role of the Laplace operator on a Riemann surface
is played by the non-linear \emph{Monge-Ampère operator} for higher
dimensional complex manifolds. A different {}``linear'' higher-dimensional
generalization of point vortex systems has previously been consider
by Kiessling \cite{ki2}, where the role of the Laplace operator is
played by the linear Paneitz operator. It involves \emph{conformal
geometry} of spheres rather than the \emph{complex (holomorphic) geometry}
considered here and the thermodynamical limit is a mean field limit
of an explicit gas with logarithmic pair interactions. 

It would be interesting to understand the relation between the present
note and the ADS/CFT correspondence \cite{agsmoo}, which relates
gravity in the bulk of a manifold to a conformal field theory on its
boundary. This is a realization of t'Hooft's holographic principle.
Such a principle has recently been put forward by Verlinde in \cite{v}
as the basis of an entropic explanation of gravity. As explained in
the concluding section \ref{sub:Conclusion-and-discussion} the emergence
of the Kähler-Einstein metric from the present microscopic model can
be interpreted as coming from a \emph{fermionic maximum entropy principle.}

From the mathematical point of view an important motivation for the
present work comes from the Yau-Tian-Donaldson program which relates
the analytic problem of existence of extremal metrics in a given Kähler
class (i.e. Kähler-Einstein metrics in the case of the canonical class)
to algebro-geometric stability conditions (notably $K-$stability;
see \cite{do1,don1,t,p-s} and references therein). For example, the
free energy functional derived below turns out to coincide, in the
canonical case, with Mabuchi's $K-$energy, which is usually used
to define various notions of $K-$stability. Moreover, the thermodynamical
convergence towards a Kähler-Einstein volume form in section \ref{sub:Convergence-in-the thermo}
is somewhat {}``dual'' to the convergence of canonically balanced
metrics conjectured by Donaldson in \cite{don2} and proved in \cite{bbgz}
(see section \ref{sub:Duality-and-relation}). 

As a conclusion one of the mathematical aims of the present paper
is to introduce a {}``thermodynamical formalism'' for Kähler-Einstein
metrics and more generally for Monge-Ampère equations of mean field
type. A rigorous mathematical account of the corresponding variational
calculus is given in \cite{berm6}.

\subsection*{Acknowledgment}

Thanks to Kurt Johansson for stimulating my interest in general $\beta-$ensembles
and to Bengt Nilsson for comments on a draft of the present paper.

\subsection{Geometric setup}

Let $X$ be a compact Kähler manifold with $\dim_{\C}X=n.$ In other
words, we are given a real manifold $(X,J)$ of dimension $D=2n$
equipped with an integrable complex structure $J$ and admitting an
Hermitian metric \[
\omega=\frac{i}{2}h_{ij}dz^{i}\wedge d\bar{z}^{j}\]
 on the complex tangent bundle $TX,$ which is closed: $d\omega=0$
(so that $\omega$ is a symplectic form) where $\omega$ is called
the Kähler form (metric). Identifying $\omega$ with a Riemannian
metric $g$ compatible with $J,$ i.e. $g=\omega(\cdot,J\cdot)$ (or
locally $g=\mbox{Re \ensuremath{h)}}$ the vacuum Einstein equations,
in Euclidean signature, with a cosmological constant read: \begin{equation}
\mbox{Ric \ensuremath{\omega=\Lambda\omega}}\label{eq:einsteins eq}\end{equation}
when $n>1$ and for general $n$ this is the equation for a Kähler-Einstein
metric. After a scaling, we may assume that the cosmological constant
$\Lambda$ is $0,1$ or $-1.$ In the following we will be mainly
concerned with the latter case, i.e. when the solution $\omega$ is
a Kähler metric with constant \emph{negative} Ricci curvature. As
shown in the seminal works of Aubin \cite{au} and Yau \cite{y} such
a metric $\omega$ exists precisely when the first Chern class $c_{1}(K_{X})$
of the canonical line bundle $K_{X}:=\Lambda^{n}(T^{*}X)$ is \emph{positive,
}which will henceforth be assumed. The Kähler-Einstein metric $\omega$
is then uniquely determined by the complex structure $J$ and we will
denote it by $\omega_{KE}.$ When $n=1,$ i.e. $X$ is a Riemann surface,
this amounts to the classical fact that $X$ admits a metric of constant
negative curvature precisely when $X$ has genus at least two. This
hyperbolic metric is unique in its conformal class (determined by
the complex structure $J)$

The starting point of the existence proof of Aubin and Yau is the
basic complex geometric fact that the metric $\omega_{KE}$ is uniquely
determined by its volume form $\omega_{KE}^{n}/n!,$ that we will
normalize to become a probability measure: \[
\mu_{KE}:=\frac{\omega_{KE}^{n}/n!}{Vn!}\]
 In other words the tensor equation \ref{eq:einsteins eq} reduces
to a\emph{ scalar} equation (for the density of $\mu_{KE})$ and the
Kähler-Einstein metric $\omega_{KE}$ may then be recovered by \[
\omega_{KE}=\frac{i}{2\pi}\partial\bar{\partial}\log\mu_{KE}\]
i.e. as $\frac{i}{2\pi}$ times the curvature two form of the metric
on the canonical line bundle $K_{X}$ defined by $\mu_{KE}.$ 

The question raised in the introduction may now be reformulated as
\emph{{}``Can the probability measure $\mu_{KE}$ be realized as
the (macroscopic) expected distribution of particles in a thermodynamical
limit of a (microscopic) statistical mechanical system canonically
associated to $X?$ }Moreover, the point is to be able to define the
microscopic system without specifying any background metric structure
so that the Einstein metric and hence (Euclidean) gravity would emerge
macroscopically. It turns out that such a statistical mechanical system
can indeed be realized by a certain free fermion gas on $X,$ as explained
below.

\subsection{General statistical mechanics formalism}

We start by recalling some basic statistical mechanical formalism.
Mathematically a (classical) gas of $N$ identical particles (i.e.
a \emph{random point process with $N$ particles}) is described by
a \emph{symmetric} probability measure $\mu^{(N)}$ on the $N-$fold
product $X^{N}$\emph{ (the $N-$particle configuration space).} In
local holomorphic coordinates $Z=(z_{1},....,z_{n})$ on the complex
manifold $X$ this means that \[
\mu^{(N)}=\rho^{(N)}(Z_{1},...,Z_{N})dV(Z_{1})\wedge\cdots\wedge dV(Z_{N})\]
 where $dV(Z_{1}):=(\frac{i}{2})^{n}dz_{1}\wedge d\bar{z}_{1}\wedge\cdots\wedge dz_{n}\wedge d\bar{z}_{n}$
and where the local \emph{$N-$point correlation function} $\rho^{(N)}$
is invariant under permutations of the $Z_{i}:$s (note that the fact
that we have not written out any dependence on $\bar{Z}_{i}$ does
not indicate that the objects are holomorphic!). Pushing forward $\mu^{(N)}$
to $X^{j}$ one then obtains the corresponding $j-$point correlation
measures $\mu_{j}^{(N)}$ on $X^{j}$ and their local densities $\rho_{j}^{(N)}.$
We will be mainly concerned with the\emph{ one-point correlation measure}
$\mu_{1}^{(N)}$ on $X,$ i.e. \[
\mu_{1}^{(N)}:=\int_{X^{N-1}}\mu^{(N)}\]
In other words, its local density $\rho_{1}^{(N)}(Z)$ represents
the probability of finding a particle in the infinitesimal box $dV(Z_{1}).$
Yet another (trivially) equivalent formulation representation of $\mu_{1}^{(N)}$
can be given: \[
\mu_{1}^{(N)}=\left\langle \frac{1}{N}\sum_{i}\delta_{x_{i}}\right\rangle ,\]
 where the brackets denote the ensemble mean (expectation) of the
random variable \begin{equation}
(x_{1},...,x_{N})\mapsto\frac{1}{N}\sum_{i}\delta_{x_{i}}\label{eq:empirical measure}\end{equation}
 with values in the space $\mathcal{M}_{1}(X)$ of probability measures
on $X.$ In other words, if $\phi$ denotes a fixed smooth function
then \[
\int_{X}\phi\mu_{1}^{(N)}=\frac{1}{N}\sum_{i}\left\langle \phi(x_{i})\right\rangle =\left\langle \phi(x_{1})\right\rangle \]
We will next explain how to define $\mu^{(N)}$ so that the one-point
correlation measures converge to the normalized volume form of the
Kähler-Einstein metric: \[
\mu_{1}^{(N)}\rightarrow\mu_{KE}\]
 in the large $N-$limit. More precisely, the convergence will hold
in the weak topology on $\mathcal{M}_{1}(X),$ i.e. \[
\int_{X}\phi\mu_{1}^{(N)}\rightarrow\int\phi\mu_{KE}\]
 for any fixed smooth function $\phi$ on $X.$ This convergence can
be interpreted as an answer to the question raised above. In fact,
the argument will give a much stronger {}``exponential'' convergence
which in particular implies the asymptotic factorization of all $j-$point
correlation functions (i.e. \emph{propagation of chaos} holds).

\subsection{Line bundles and Slater determinants}

To define $\mu^{(N)}$ first recall that we have assumed that the
canonical line bundle $K_{X}\rightarrow X$ is\emph{ positive} (i.e.
ample in the sense of algebraic geometry). We next recall some basic
facts about line bundles (see for example \cite{b-v-,d-k} for introductions
aimed at physicists). To any holomorphic line bundle $L\rightarrow X$
there is a naturally associated $N-$dimensional complex vector space
$H^{0}(X,L)$ consisting of global holomorphic section of $L\rightarrow X$
and the limit we will be interested is when $L$ is replaced by a
large tensor power $L^{\otimes k}.$ Since $L$ is assumed ample it
follows that the dimension $N=N_{k}$ (which will be the number of
particles of our gas) grows with $k$ in the following way: \[
N_{k}:=\dim_{\C}H^{0}(X,L^{\otimes k})=Vk^{n}+o(k^{n})\]
 where the \emph{volume} $V>0.$ In particular, 

\[
N(=N_{k})\rightarrow\infty\Leftrightarrow k\rightarrow\infty\]
We will often omit the subscript $k$ in $N_{k}$.

In physics, $H^{0}(X,L)$ usually arises as the quantum ground state
space of a single \emph{chiral fermion} on $X$ coupled to $L$ \cite{vv,b-v-},
since it may be realized as the zero of a gauged Dirac operator (once
metrics/gauge fields are introduce, as explained below). The corresponding
$N-$particle space of fermions is then, according to Pauli's exclusion
principle, represented by the top exterior power $\Lambda^{N}H^{0}(X,L).$
In other words this is the maximally filled many particle fermion
state. As a consequence it is one-dimensional and may, up to scaling,
be represented by the $N-$body state \[
\Psi(x_{1},...x_{N}):=\Psi_{1}(x_{1})\wedge\cdots\wedge\Psi_{N}(x_{N})\]
 expressed in terms of a given base $(\Psi_{I})$ in $H^{0}(X,L),$
where $I=1,...,N.$ Locally this means that $\Psi$ may be written
as a Slater determinant: \begin{equation}
\Psi(Z_{1},...,Z_{N})=\det(\Psi_{I}(Z_{J}))\label{eq:slater det}\end{equation}
 which hence transforms as a holomorphic section of the line bundle
$L^{\boxtimes N}$ over $X^{N}.$

\subsubsection{Introducing metrics}

Usually, one equips $L$ with an\emph{ Hermitian metric} $h_{0}.$
Taking the squared point-wise norm $\left\Vert \Psi(Z)\right\Vert ^{2}$
with respect to $h_{0}$ of a section $\Psi$ of $L$ hence gives
a scalar function on $X.$ Let us briefly recall the notion of curvature
in this context. The (Chern)\emph{ curvature form }$\Theta$ of $h_{0}$
is the globally well-defined two-form on $X$ locally defined as follows:
if $s$ is a local trivializing holomorphic section of $L,$ then
\begin{equation}
\Theta:=-\partial\bar{\partial}\log(\left\Vert s\right\Vert ^{2})\label{eq:def of curvature}\end{equation}
 Physically, the curvature form $\Theta$ represents a background
\emph{magnetic }two-form of bidegree $(1,1)$ to which the fermions
are minimally coupled. More precisely, the holomorphic structure on
$L$ together with the Hermitian metric $h_{0}$ determines a reduction
of $L$ to a $U(1)-$bundle and a unique unitary connection $A$ on
$L,$ i.e. a $U(1)-$gauge field such that its field strength $F_{A}=\Theta$
is of type $(1,1)$ \cite{b-v-,d-k}. Let us briefly recall the local
meaning of this correspondence. In terms of a local holomorphic trivialization
$s$ of $L$ we may write \[
\Psi(z)=f(z)s\]
 where $f(z)$ is a local holomorphic function, i.e. $\bar{\partial}f=0.$
Writing $h_{0}(z):=\left\Vert s\right\Vert ^{2}=e^{-\Phi}$ for a
local function $\Phi$ we then have \[
\left\Vert \Psi(Z)\right\Vert ^{2}=|f(z)|^{2}e^{-\Phi}=|f_{\Phi}(z)|^{2}\]
where $f_{\Phi}:=f(z)e^{-\Phi/2}$ represents $\Psi$ wrt the unitary
(but non-holomorphic) local trivialization $e^{\Phi/2}s.$ Note in
particular that \begin{equation}
\bar{\partial}_{\Phi}f_{\Phi}=0,\,\,\,\,\,\,\bar{\partial}_{\Phi}=\bar{\partial}+\frac{1}{2}\bar{\partial}\Phi.\label{eq:local dirac equation}\end{equation}
Hence, setting \[
A=\frac{1}{2}(\bar{\partial}\Phi-\partial\Phi),\,\,\, F_{A}:=dA=\partial\bar{\partial}\Phi\]
gives a one-form with values in $i\R$ representing the $U(1)-$gauge
field (connection) and the equation \ref{eq:local dirac equation}
implies that $f_{\Phi}$ is a zero-mode for the Dirac operator $\D_{A}$
expressed in the unitary trivialization (where $\D_{A}$ also depends
on a choice of metric on $X;$ compare section \ref{sub:Derivation-of-}
below). 

The metric $h_{0}$ is \emph{positively curved }precisely when the
\emph{real} two-form \begin{equation}
\omega:=\frac{i}{2\pi}\Theta(=\frac{i}{2\pi}\partial\bar{\partial}\Phi)\label{eq:def of norm curv}\end{equation}
is positive definite, i.e. when it defines a \emph{Kähler metric}
on $X$ (and $\Phi$ is sometimes called the local Kähler potential
of $\omega$). The line bundle $L$ is ample precisely when it admits
some positively curved metric. The normalization above ensures that
the cohomology class $[\omega],$ which represents the normalized
\emph{first Chern class} $c_{1}(L)$ is an integer class, i.e. it
lies in the integer lattice $H^{2}(X,\Z)$ of $H^{2}(X,\R).$ 

The Hermitian metric $h_{0}$ naturally induces metrics on all tensor
powers of $L$ etc. Coming back to the Slater determinant above, the
point-wise squared norm with respect to the metric $h_{0}$ \[
\left\Vert \Psi(Z_{1},...,Z_{N})\right\Vert ^{2}\]
 is, from a physical point of view, proportional to the probability
of finding (or creating) particles at the point $Z_{1},...$$Z_{N}$
on $X$ in the presence of the corresponding background magnetic field.
To normalize it we also need to pick an integration measure $\mu_{0}$
(a volume form) on $X$ so that \[
\left\Vert \Psi(Z_{1},...,Z_{N})\right\Vert ^{2}/\mathcal{Z}_{N},\,\,\,\mathcal{Z}_{N}:=\int_{X^{N}}\left\Vert \Psi\right\Vert ^{2}\mu_{0}^{\otimes N}\]
is a probability density on $X^{N}.$ Since, $\Lambda^{N}H^{0}(X,L)$
is one-dimensional the probability density above is, by homogeneity,
independent of the choice of base $(\Psi_{I})$ in $H^{0}(X,L),$
but, of course, it does depend on the metric $h_{0}$ on $L$ (or
more precisely on the background magnetic field $F_{A}$) and also
on the integration measure $\mu_{0}$ on $X.$

\subsection{The canonical background free ensemble \label{sub:The-canonical-background}}

The main point of the present paper is the simple observation that
in the particular case when $L$ is the canonical line bundle $K_{X}$
there is no need to specify any metric on $K_{X}$ if one defines
a probability measure on $X^{N}$ by \[
\mu^{(N)}=(\Psi_{1}\wedge\bar{\Psi}_{1}\wedge\cdots\wedge\Psi_{N}\wedge\bar{\Psi}_{N})^{1/k}/\mathcal{Z_{N}}.\]
To see this first note that it follows from the very definition of
$K_{X}$ that $(\Psi_{1}\wedge\bar{\Psi}_{1}\wedge\cdots\wedge\Psi_{N}\wedge\bar{\Psi}_{N})^{1/k}$
transforms as a (degenerate) volume form on $X^{N}.$ The points is
that if $\Psi$ is a section of $K_{X}(:=\Lambda^{n}(TX))\rightarrow X$
then $\Psi\otimes\bar{\Psi}$ gives a well-defined measure on $X$
(i.e. a degenerate volume form), concretely this is a consequence
of the local representation $\Psi=f(z_{1},...,z_{N})dz_{1}\wedge\cdots\wedge dz_{N}.$
Hence after dividing by \[
\mathcal{Z_{N}}=\int_{X^{N}}(\Psi_{1}\wedge\bar{\Psi}_{1}\wedge\cdots\wedge\Psi_{N}\wedge\bar{\Psi}_{N})^{1/k}=\int_{X^{N}}|\det(f_{I}(Z_{J}))|^{2/k}dV(Z_{1})\wedge\cdots\wedge dV(Z_{N})\]
 one obtains a probability measure $\mu^{(N)}$ on $X^{N_{k}}$ which
is canonically associated to $(X,K_{X}^{\otimes k}),$ since by homogeneity
it is independent of the base $(\Psi_{I})$ in $H^{0}(X,K_{X}^{\otimes k}).$
Note that when $n=1,$ i.e. $X$ is a Riemann surface of genus at
least two the space $H^{0}(X,K_{X}^{\otimes k})$ arises as the space
of spin $2k$ particles \cite{vv,b-v-} .

\subsection{\label{sub:General-ensembles.}General $\beta-$ensembles.}

Before turning to the investigation of the thermodynamical convergence
towards the Kähler-Einstein volume form $\mu_{KE}$ it should be pointed
out that\emph{ integer} powers of Slater determinants have been used
before to model the \emph{fractional Quantum Hall effect} \cite{l}.
More generally we note that the previous construction may be generalized
by introducing general $k-$dependent powers $\beta_{k}$ in the Slater
determinant. To see this we come back to the general setting of an
ample line bundle $L\rightarrow X$ and now fix a background metric
$h_{0}$ on $L$ and a volume form $\mu_{0}$ on $X.$ To this geometric
data we associate the probability measure \[
\mu^{(N_{k})}=\left\Vert \Psi\right\Vert ^{\beta_{k}}\mu_{0}^{\otimes N}/\mathcal{Z_{N}}.\]
on $X^{N}$ for a fixed choice of parameters $\beta_{k}.$ The case
of $L=K_{X}$ considered above is obtained by setting $\beta_{k}=2/k,$
fixing any metric $h_{0}$ on $K_{X}$ and then letting $\mu_{0}=1/h_{0},$
which defines a volume form on $X.$ Then it is easy to see that all
factors of $h_{0}$ cancel out leading to the previous canonical construction
above. Finally, note that if one defines the\emph{ Hamiltonian} \[
H^{(N)}:=-\log\left\Vert \Psi\right\Vert \]
 then $\mu^{(N_{k})}$ may be represented as a Boltzmann-Gibbs ensemble
\begin{equation}
\mu^{(N_{k})}=e^{-\beta_{k}H^{(N)}}\mu_{0}^{\otimes N}/\mathcal{Z_{N}},\label{eq:bolt-gibbs}\end{equation}
 of a \emph{classical }system in thermal equilibrium with an external
heat bath of temperature $T_{k}=1/\beta_{k}.$ From this point of
view $\mathcal{Z_{N}}$ is the \emph{partition function} of the system.
It depends of the choice of bases $(\Psi_{I})$ in $H^{0}(X,L^{\otimes k})$
(but $\mu^{(N_{k})}$ does not, as explained above). For example,
the case when $\beta_{k}=1,$ $2$ or $4$ appears in the study of
the Random Matrix ensembles associated to the classical groups (see
\cite{j} and references therein). In this latter cases $X$ is taken
as the Riemann sphere, i.e. the compactification of $\C$ and then
the Hilbert space $H^{0}(X,L^{\otimes k})$ appears as the corresponding
orthogonal polynomials on $\C.$ However, as will be explained below
we will be different in the case when $\beta_{k}$ depends on $k$
and is of the order $1/k$ which changes the thermodynamical limit,
as compared to ordinary Random Matrix Theory.

It is worth emphasizing that the Hamiltonian $H^{(N)}$ above is not
a sum of pair interactions (even to the leading order) when $n>1.$
This is closely related to the fact that the mean field equations
obtained in section \ref{sub:The-minimizer-} are fully non-linear
(i.e. non-linear in the derivative terms, so that the corresponding
actions are of higher order) and it makes the analysis of the thermodynamical
limit rather challenging.

\section{\label{sub:Convergence-in-the thermo}Convergence in the thermodynamical
limit}

It will be illuminating to consider the general setting of the previous
setting with \[
\beta_{k}=\beta/k\]
 for a fixed parameter $\beta$ (where $\beta=2$ appears in the canonical
background free case \ref{sub:The-canonical-background}). As will
be clear this is, in a certain sense, a mean field limit. As explain
above we hence fix the geometric data $(h_{0},\mu_{0})$ consisting
of Hermitian metric $h_{0}$ on $L\rightarrow X$ and a volume form
$\mu_{0}$ on $X$ (the canonical case is then a special case if one
takes $h_{0}=1/\mu_{0}$ for any given volume form $\mu_{0}).$ Given
this data we furthermore fix a base $(\Psi_{I})$ in $H^{0}(X,L^{\otimes k}),$
for any $k,$ which is orthonormal with respect to Hilbert space structure
on $H^{0}(X,L^{\otimes k})$ induced by $(h_{0},\mu_{0}):$ \[
\left\langle f,g\right\rangle _{X}:=\int_{X}\left\langle f,g\right\rangle \mu_{0},\]
where the point-wise Hermitian product in the integrand is taken with
respect $h_{0}.$ In particular, the corresponding partition function
is then (a power of) the induced $L^{\beta/k}$ norm of the corresponding
Slater determinant $\Psi$ (formula \ref{eq:slater det}): \[
\mathcal{Z}_{N}:=\int_{X^{N}}\left\Vert \Psi\right\Vert ^{\beta/k}\mu_{0}^{\otimes N}\]
 To prove the convergence we will use the techniques of the \emph{theory
of large deviations. }In a nutshell this is a formalism which allows
one to give a meaning to the statement that a given sequence of probability
measure $\mu^{(N)}$on $X^{N}$ is {}``exponentially concentrated
on a deterministic macroscopic measure $\mu_{*}$ with a rate functional
$I(\mu)"$ (see \cite{to} for an introduction to the theory of large
deviations, due to Cramér, Varadhan and others, emphasizing the links
to statistical mechanics - relations to functional integrals are emphasized
in \cite{gaw}) . Heuristically, the idea is to think of the large
$N-$limit of the $N-$particle space $X^{N}$ of {}``microstates''
as being approximated by a space of {}``macrostates'', which is
the space $\mathcal{M}_{1}(X)$ of all probability measures on $X:$
\[
X^{N}\sim\mathcal{M}_{1}(X),\]
 as $N\rightarrow\infty.$ The exponential concentration referred
to above may then be heuristically written as \begin{equation}
\mu^{(N)}:=\rho^{(N)}(Z_{1},...,Z_{N})dV(Z_{1})\wedge\cdots dV(Z_{N})\sim e^{-NF(\mu)}\mathcal{D}\mu,\label{eq:large dev heur}\end{equation}
 where $\mathcal{D}\mu$ denotes a (formal) probability measure on
the infinite dimensional space $\mathcal{M}_{1}(X)$ (more generally,
the exponent $N$ could be replaced by a \emph{rate} $a_{N}$ which
is usually a power of $N),$ where $F(\mu)$ is called the \emph{rate
functional.} Exponential concentration around $\mu_{*}$ appears when
$F(\mu)\geq0$ with $\mu_{*}$ the unique minimizer of $F.$ Mathematically,
the {}``change of variables'' from $X^{N}$ to $\mathcal{M}_{1}(X)$
is made precise by using the embedding \[
j_{N}:\, X^{N}\rightarrow\mathcal{M}_{1}(X),\,\,\, j_{N}(x,...,x_{N}):=\frac{1}{N}\sum_{i}\delta_{x_{i}}\]
and then pushing forward $\mu^{(N)}$ to $\mathcal{M}_{1}(X)$ with
the map $j_{N},$ giving a probability measure $(j_{N})_{*}\mu^{(N)}$
on $\mathcal{M}_{1}(X)$ (i.e. the law of the random variable \ref{eq:empirical measure}).
The precise mathematical meaning of \ref{eq:large dev heur}, in the
sense of large deviations, is then that 

\[
\lim_{\delta\rightarrow0}\lim_{N\rightarrow\infty}\frac{1}{N}\log\int_{\mathcal{B}_{\delta}(\mu)}(j_{N})_{*}\mu^{(N)}=-F(\mu),\]
 integrating over a small ball $\mathcal{B}_{\delta}(\mu)$ of radius
$\delta$ centered at $\mu\in\mathcal{M}_{1}(X)$ (using any metric
on $\mathcal{M}_{1}(X)$ which is compatible with the weak topology)
and where we have assumed for simplicity that $F$ is continuous (however
in our case the definition of the LDP is slightly more complicated
as $F$ will only be semi-continuous). 

The idea is now to establish the asymptotics \ref{eq:large dev heur}
for a certain\emph{ free energy functional} $F(\mu)$ which is minimized
precisely on a measure $\mu_{*}$ which equals the Kähler-Einstein
measure $\mu_{KE}$ in the canonical case introduced in section \ref{sub:The-canonical-background}.
In fact, in this latter case the functional $F(\mu)$ will turn out
to be naturally identified with Mabuchi's K-energy, which plays an
important role in Kähler geometry (as explained in section \ref{sub:Duality-and-relation})

To this end we will combine two already established asymptotics, concerning
the the case when $\beta_{k}=0$ and $\beta_{k}=2$ respectively.
In the first case it is a classical result going back to the work
of Boltzmann (called Sanov's theorem in the mathematics literature)
that the asymptotics \ref{eq:large dev heur} hold with $-F(\mu)$
equal to the relative entropy functional $S(\mu)$: \begin{equation}
\mu_{0}^{\otimes N}\sim e^{NS(\mu)}\mathcal{D}\mu,\label{eq:large dev as beta zero}\end{equation}
 where \begin{equation}
S(\mu):=-\int_{X}\log(\frac{\mu}{\mu_{0}})\mu(\leq0)\label{eq:def of entropy}\end{equation}
 if $\mu$ has a density with respect to $\mu_{0}$ and otherwise
$S(\mu)=-\infty.$ This result gives a precise meaning to Boltzmann's
notion of entropy as proportional to the logarithmic number (or volume)
of microstates corresponding to a given macrostate.

Next, in the case when $\beta_{k}=2$ it was shown very recently in
\cite{berm3} that \begin{equation}
e^{-2H^{(N)}(x_{1},...,x_{N})}\mu_{0}^{\otimes N}\sim e^{-kNE(\mu)}\mathcal{D}\mu\label{eq:large deve as beta 2}\end{equation}

In the present work we are interested in the intermediate asymptotic
regime where $\beta_{k}=\beta/k.$ Decomposing the corresponding probability
measure $\mu^{(N)}$ as 

\[
\mu^{(N)}:=(e^{-2H^{(N)}(x_{1},...,x_{N})})^{\beta/2k}\mu_{0}^{\otimes N}\]
 or more precisely as \[
\mu^{(N)}:=\left[(e^{-2H^{(N)}(x_{1},...,x_{N})}\mu_{0}^{\otimes N})^{\beta/2k}\right]\left[\cdot(\mu_{0}^{\otimes N})^{1-\beta/2k}\right]\]
we can, at least heuristically, combine the asymptotics \ref{eq:large dev as beta zero}
and \ref{eq:large deve as beta 2} to get \begin{equation}
\mu^{(N)}\sim e^{-N\beta(E(\mu)-\frac{1}{\beta}S(\mu))}\mathcal{D}\mu\label{eq:rasympt with rate fun}\end{equation}
 (in order to be mathematically rigorous this heuristic argument needs
to be complemented with precise estimates justifying the {}``interchange''
of the large $N$ and small $\delta-$limits)

The convergence of of the one-point correlation measures $\mu_{1}^{(N)}$
towards the minimizer $\mu_{*}$ of the functional \[
F(\mu):=E(\mu)-\frac{1}{\beta}S(\mu))\]
can now be shown by standard arguments (given the existence and uniqueness
of $\mu_{*}$ which we will deal with in section \ref{sub:The-minimizer-}).
First note that the partition function may be asymptotically calculated
as

\[
\mathcal{Z_{N}}\sim\int_{\mathcal{M}_{1}(X)}\mathcal{D}\mu e^{-\beta N(E(\mu)-\frac{1}{\beta}S(\mu))},\]
 giving \[
-\frac{1}{\beta N}\mathcal{\log Z_{N}}\rightarrow\inf_{\mu\in\mathcal{M}_{1}(X)}(E(\mu)-\frac{1}{\beta}S(\mu)\]
 Next, note that upon performing an overall scaling of the original
base $(\Psi_{I})$ we may assume that the infimum above vanishes.
Now fix a smooth function $\phi$ on $X$ and consider the functional
\[
\mathcal{F}_{N}(u):=-\log\left\langle e^{-(\phi(x_{1})+...+\phi(x_{N}))}\right\rangle :=-\log\int_{X^{N}}e^{-(\phi(x_{1})+...+\phi(x_{N}))}\mu^{(N_{k})}\]
The following basic general exact variational identity holds \begin{equation}
\frac{1}{\beta N}\frac{d\mathcal{F}_{N}(t\phi)}{dt}_{t=0}=\int_{X}\mu_{1}^{(N_{k})}\phi\label{eq:finite n var ident}\end{equation}
Arguing precisely as above and using the trivial asymptotics \begin{equation}
e^{-(\phi(x_{1})+...+\phi(x_{N}))}\sim e^{-N\int_{X}\phi\mu},\label{eq:exp term as mea}\end{equation}
 hence gives \[
\frac{1}{\beta N}\mathcal{F}_{N}(t\phi)\rightarrow\inf_{\mu\in\mathcal{M}_{1}(X)}(E(\mu)-\frac{1}{\beta}S(\mu)+t\int_{X}\phi\mu)\]
Finally, differentiating with respect to $t$ gives \[
\frac{1}{\beta N}\frac{d\mathcal{F}_{N}(t\phi)}{dt}_{t=0}\rightarrow0+\int_{X}\mu_{*}\phi\]
 and hence, using \ref{eq:finite n var ident}, we finally get \[
\mu_{1}^{(N)}\rightarrow\mu_{*}\]
Next, we will show that the minimizer $\mu_{*}$ can be obtained by
solving a mean field type equation which will reduce to the Kähler-Einstein
equation in the canonical case. We will start by explaining the notion
of pluricomplex energy $E(\mu)$ appearing in the asymptotics \ref{eq:large deve as beta 2}.

\subsection{The pluricomplex energy $E(\mu)$}

Assume now that the fixed metric $h_{0}$ on $L$ has positive curvature,
i.e. its normalized curvature form is a Kähler form that we denote
by $\omega(:=\omega_{0}).$ Since, $h_{0}$ is uniquely determined
up to scaling by its curvature form $\omega$ and since the probability
measure $\mu^{(N)}$ is insensitive to scaling of $h_{0}$ we may
as well say that the geometric data defining the $\beta_{k}$- ensemble
is $(\omega_{0},\mu_{0}).$ 

Now any Kähler metric which is cohomologous to $\omega$ (i.e. in
the class $[\omega]=c_{1}(L))$ may by the $\partial\bar{\partial}-$lemma
be written as \[
\omega_{\phi}:=\omega+\frac{i}{\pi}\partial\bar{\partial}\phi\]
 for a smooth function $u.$ In this way the space of all Kähler metrics
in $c_{1}(L)$ be identified with the space of (global) \emph{Kähler
potentials} \[
\mathcal{H}_{\omega}(X):=\{\phi\in\mathcal{C}^{\infty}(X):\,\omega_{\phi}>0\}\]
modulo constants (we will usually mod out by $\R$ sometimes without
mentioning it explicitly). Note that these potentials are globally
well-defined and depend on the choice of $\omega$ Geometrically,
the space $\mathcal{H}_{\omega}(X)$ may be identified with the space
$\mathcal{H}(L)$ of \emph{all positively curved Hermitian metrics
on $L$} and $\omega_{\phi}$ with the (normalized) curvature form
of the metric \[
h_{\phi}:=e^{-2\phi}h_{0}\]
 on $L$ corresponding to $\phi$ (as follows immediately from formula
\ref{eq:def of curvature}), i.e. $\phi:=\Phi-\Phi_{0}$ in terms
of \emph{local} Kähler potentials of $\omega_{\phi}$ and $\omega$
respectively. 

Thanks to Yau's solution of the Calabi conjecture one can also associate
potentials to \emph{volume forms} on $X.$ Indeed, to any volume form
$\mu$ on $X$ (which we will always assume normalized so that $\int_{X}\mu=1)$
there is a unique\emph{ potential} $\phi(:=\phi_{\mu})$ in $\mathcal{H}_{\omega}(X)/\R$
such that \begin{equation}
\frac{\omega_{\phi}^{n}}{Vn!}=\mu,\label{eq:inhomogenous ma-eq}\end{equation}
 where $V$ is the \emph{volume} of any Kähler metric in the class
$c_{1}(L).$ The equation involves the $n:$th exterior power of $\omega_{\phi}$
and is hence a non-linear generalization of the inhomogeneous Laplace
equation, called the inhomogeneous (complex) \emph{Monge-Ampère} \emph{equation}
(and the left hand side above is called the \emph{Monge-Ampère measure}
of $\phi).$ 

The previous equation can also be given a variational formulation
by noting that there is a functional $\mathcal{E}_{\omega}$ (we will
often omit the subscript $\omega)$ on the space $\mathcal{H}_{\omega}(X)$
such that its first variation is given by \begin{equation}
\delta\mathcal{E}(\phi):=d\mathcal{E}_{\phi}=\frac{\omega_{\phi}^{n}}{Vn!},\label{eq:var def of ma action}\end{equation}
 where $d\mathcal{E}$ is the differential of $\mathcal{E}(\phi)$
seen as a one-form on $\mathcal{H}_{\omega}(X).$ The functional $\mathcal{E}$
is uniquely determined by the normalization $\mathcal{E}(0)=0$ (singled
out by the fixed reference Kähler metric $\omega).$ This is a well-known
functional in Kähler geometry which seems to first have been introduced
by Mabuchi ( (it is denoted by $-F_{\omega}$ in the book \cite{t};
similar functionals also appeared in the works of Aubin and Yau).
We will call it the\emph{ Monge-Ampère action}, since it physically
appears as an action generalizing the\emph{ Liouville action}, as
explained in section \ref{sub:Derivation-of-}. It is straight-forward
to obtain an explicit formula for $\mathcal{E}(\phi)$ by integrating
along the line segment $t\phi$ for $0\leq t\leq1$ and get \begin{equation}
\mathcal{E}_{\omega}(\phi)=\frac{1}{(n+1)!V}\int_{X}\phi\sum_{i=1}^{n}(\omega^{n-j}\wedge\omega_{\phi}^{j}),\label{eq:explcit ex for e of phi}\end{equation}
 but we will only make use of the defining property \ref{eq:var def of ma action}
in the following.

The functional $\mathcal{E}$ is (strictly) concave on $\mathcal{H}_{\omega}(X)/\R$
(with respect to the flat metric) \cite{t} and hence the potential
$\phi_{\mu}$ may be characterized as the unique (mod $\R)$ maximizer
of the the functional \[
\phi\mapsto\mathcal{E}(\phi)-\left\langle \phi,\mu\right\rangle ,\]
 expressed in terms of the usual pairing \begin{equation}
\left\langle \phi,\mu\right\rangle :=\int_{X}\phi\mu\label{eq:pairing}\end{equation}
Finally, we can now, following \cite{bbgz}, define the \emph{(pluricomplex)
energy} $E_{\omega}(\mu)$ of the measure $\mu$ (but we will often
omit the subscript $\omega)$ as \[
E(\mu):=\sup_{\phi\in\mathcal{H}_{\omega}(X)}\mathcal{E}(\phi)-\left\langle \phi,\mu\right\rangle =\mathcal{E}(\phi_{\mu})-\left\langle \phi_{\mu},\mu\right\rangle \]
The first equality in fact makes sense for \emph{any }(possibly singular)
measure $\mu$ in $\mathcal{M}_{1}(X)$ and one says that $\mu$ has
finite energy if $E(\mu)<\infty.$ 

When $n=1$ one may actually take the sup defining $E$ over \emph{all}
$\phi\in\mathcal{C}^{\infty}(X)$ (i.e. without imposing the constraint
$\omega_{\phi}>0).$ Then the convex functional $E(\mu)$ is, by definition,
the Legendre transform of the concave functional $\mathcal{E}(\phi)$
on $\mathcal{C}^{\infty}(X)$ (with a non-standard sign convention).
It turns out that in the case $n>1$ the functional $E(\mu)$ can
also be realized as a Legendre transform by extending $\mathcal{E}(\phi)$
to another (concave and one time differentiable) functional $\mathcal{F}_{\infty}$
on $\mathcal{C}^{\infty}(X)$ \cite{bbgz,berm3,b-b} (which appears
in the general asymptotics \ref{eq:conv of free energy det proc}
below). This fact is an important ingredient in the variational approach
to complex Monge-Ampère equations introduced in \cite{bbgz}. 

Note that the energy functional $E$ certainly depends on the choice
of fixed Kähler metric $\omega.$ In fact, it is not hard to see that
$E(\mu)\geq0$ with equality precisely if $\mu=\omega^{n}/Vn!.$ Indeed,
it follows from general principles (concerning Legendre transforms)
that \[
\inf_{\mu\in\mathcal{M}_{1}(X)}E(\mu)=E((\delta\mathcal{E})(0))=E(\frac{\omega^{n}}{Vn!})=\mathcal{E}(0)-0=0\]
It would hence be more appropriate to call $E(\mu)$ the\emph{ relative
energy of $\mu.$}

\subsubsection{The Riemann surface case}

It may be illuminating to consider the case when $n=1,$ i.e. when
$X$ is a Riemann surface. Then $\mathcal{E}(\phi)$ coincides with
the functional sometimes referred to as the \emph{Liouville action}
in the physics literature \cite{b-v-,o-v}: \[
\mathcal{E}(\phi)=\frac{1}{2}\int_{X}\phi(\omega_{\phi}+\omega)\]
 and hence, taking the potential $\phi_{\mu}$ to be normalized so
that $\int\phi_{\mu}\omega=0$ we get \begin{equation}
E(\mu)=-\frac{i}{2\pi}\int_{X}\phi_{\mu}\partial\bar{\partial}\phi_{\mu}=\frac{i}{2\pi}\int\partial\phi_{\mu}\wedge\bar{\partial}\phi_{\mu}\label{eq:energy as gradient}\end{equation}
 which is essentially the usual electrostatic energy of the continuous
charge distribution $\mu$ in the neutralizing background charge $-\omega.$
Equivalently, if we define the Green function $g(x,y)$ for the scalar
Laplacian $\Delta=\omega^{-1}\frac{i}{\pi}\partial\bar{\partial}$
on $X$ by the properties $g(x,y)=g(y,x)$ and \[
\frac{i}{\pi}\partial_{x}\bar{\partial}_{x}g(x,y)=\delta_{x}(y)-\omega(y),\,\,\,\int_{X}g(x,y)\omega(y)=0\]
then we have $\phi_{\mu}(x)=\int_{X}g(x,y)d\mu(y)$ and hence \[
E(\mu)=-\frac{1}{2}\int_{X\times X}g(x,y)d\mu(x)\otimes d\mu(y).\]

\subsection{\label{sub:The-minimizer-}The minimizer $\mu_{*}$ of the free energy
functional $F(\mu)$ and mean field equations}

In this section we will give some formal variational arguments to
determine the minimizer of $F(\mu).$ See \cite{berm6} for a rigorous
account and further developments. Recall that the free energy functional
$F(\mu)$ (for a fixed parameter $\beta>0)$ on the space $\mathcal{M}_{1}(X)$
of probability measure son $X$ is defined by \[
F(\mu):=E(\mu)-\frac{1}{\beta}S(\mu)\]
where $E$ is the energy functional define in the previous section
and $S(\mu)$ is the relative entropy \ref{eq:def of entropy}. It
follows from basic duality arguments that $F$ is strictly convex
on $\mathcal{M}_{1}(X)$ (or rather on the subset where $F$ is finite)
and hence admits at most one minimizer. Next we note that $\mu$ is
a critical point for $F(\mu)$ on $\mathcal{M}_{1}(X)$ if and only
if \[
-\phi_{\mu}+\frac{1}{\beta}\log(\mu/\mu_{0})-Z_{\mu}=0,\]
 where $\phi_{\mu}$ is the potential of $\mu$ and $Z_{\mu}$ is
a normalizing constant. Indeed, using the defining properties of $\mathcal{E}$
and $E$ respectively one obtains (by basic Legendre transform considerations)
that \[
\delta E(\mu)=-\phi_{\mu}\]
as a one-form on the infinite dimensional submanifold $\mathcal{M}_{1}(X)$
of the vector space $\mathcal{M}(X)$ of all signed measures. Moreover,
a simple calculation gives \[
\delta S(\mu)=-\log(\mu/\mu_{0})+Z_{\mu},\]
 where $Z_{\mu}$ is a normalizing constant (coming from the constraint
$\int_{X}\mu=1).$ Combining these two variational formulas gives
\begin{equation}
\delta F(\mu)=-\log(\mu/\mu_{0})-\phi_{\mu}\label{eq:variational der of f}\end{equation}
up to a normalizing constant. In other words, $\mu$ is a critical
point for for $F(\mu)$ on $\mathcal{M}_{1}(X)$ if and only if its
potential $\phi$ solves the following non-linear partial differential
equation: \begin{equation}
\frac{\omega_{\phi}^{n}}{Vn!}=\frac{e^{\beta\phi}\mu_{0}}{Z_{\phi}}\label{eq:mean field equation}\end{equation}
 As follows from a simple modification of the proof of the Aubin-Yau
theorem there is a unique $\phi\in\mathcal{H}_{\omega}(X)/\R$ solving
this equation (crucially using that $\beta>0)$ which by strict convexity
is hence the unique maximizer of the free energy functional $F.$
It is sometimes convenient to fix the normalization of the solution
$\phi$ above by imposing that \[
\int_{X}e^{\beta\phi}\mu_{0}=1,\]
 i.e. $\phi\in\mathcal{H}_{\omega}(X)$ is the \emph{unique} solution
to \begin{equation}
\frac{\omega_{\phi}^{n}}{Vn!}=e^{\beta\phi}\mu_{0}\label{eq:m-a mfe ej z}\end{equation}
It should be pointed out that when $n=1$ the previous equation is
often called the \emph{mean field equation} \cite{clmp,k} and accordingly
we will call it the\emph{ mean field Monge-Ampère equation} for a
general dimension $n.$ 

Finally, coming back to the canonical case when $\beta=2$ and $L=K_{X}$
we take, as explained in section \ref{sub:The-canonical-background}
the geometric data $(\omega,\mu_{0})$ such that $\omega$ is the
curvature form of the metric on $K_{X}$ defined by the inverse $1/\mu_{0}.$
This means that $\mu_{0}=e^{2f_{\omega}}\omega^{n}/Vn!,$ where $f_{\omega}$
is the \emph{Ricci potential, i.e.} \[
\frac{i}{\pi}\partial\bar{\partial}f_{\omega}=\omega+\mbox{Ric\ensuremath{\omega},\,\,\,\ensuremath{\int_{X}e^{2f_{\omega}}\omega^{n}/Vn!=1}},\]
 where $\mbox{Ric}\omega=-\frac{i}{\pi}\partial\bar{\partial}(\log\omega^{n})$
is the $(1,1)-$form representing the Ricci curvature of $\omega.$
Then the corresponding Monge-Ampère mean field equation reads \[
\omega_{\phi}^{n}=e^{2\phi}e^{2f_{\omega}}\omega^{n}\]
Hence, the solution $\phi$ is such that the Kähler metric $\omega_{\phi}$
satisfies \[
\mbox{Ric\ensuremath{\omega_{\phi}=-\omega_{\phi}}},\]
 i.e. $\omega_{\phi}$ is a Kähler-Einstein metric with negative Ricci
curvature. Coming back to the convergence of the one-correlation measures
in the thermodynamical limit, considered in section \ref{sub:Convergence-in-the thermo},
this means that the limiting measure $\mu_{*}$ indeed equals $\mu_{KE}:=\omega_{KE}^{n}/Vn!,$
as expected.

\subsubsection{\label{sub:Derivation-of-}The asymptotics \ref{eq:large deve as beta 2}
for $\beta_{k}=2$ and effective bosonization}

Before continuing let us briefly explain the idea behind the large
deviation asymptotics \ref{eq:large deve as beta 2} proved in \cite{berm3}
and its relation to effective bosonization. The starting point is
the basic observation that when $\beta_{k}=2$ the one point correlation
function $\rho_{1}^{(N)}$ can be represented as a \emph{density of
states function:} \[
\rho_{1}^{(N)}(Z)=\sum_{I=1}^{N}\left\Vert \Psi_{I}(Z)\right\Vert ^{2}\]
(called the \emph{Bergman kernel at the diagonal} in the mathematics
literature). By a fundamental result of Bouche and Tian the leading
asymptotics of the corresponding one point correlation measure are
given by the Monge-Ampère measure: \[
\mu_{1}^{(N)}\rightarrow\omega^{n}/Vn!\]
(see \cite{z} for a survey of Bergman kernel asymptotics and \cite{d-k}
for a physical point of view) Now using these asymptotics and perturbing
by potentials $\phi$ in $\mathcal{H}_{\omega}(X)$ one can reverse
the arguments used in the end of section \ref{sub:Convergence-in-the thermo}
and get \[
\frac{1}{N}\mathcal{F}_{N}(\phi)\rightarrow\mathcal{E}(\phi),\,\,\,\phi\in\mathcal{H}_{\omega}(X)\]
 using the variational property of $\mathcal{E}.$ Then an argument
involving Legendre transforms gives the large deviation asymptotics
\ref{eq:large deve as beta 2}, using that $E(\mu)$ can be realized
as an (infinite dimensional) Legendre transform of $\mathcal{E}(\phi).$
More precisely, the argument uses the convergence of (perturbed) free
energies \begin{equation}
\frac{1}{N}\mathcal{F}_{N}(\phi)\rightarrow\mathcal{F}_{\infty}(\phi),\,\,\,\phi\in\mathcal{C}^{\infty}(X)\label{eq:conv of free energy det proc}\end{equation}
 for\emph{ any} smooth function $\phi$ (not necessarily with $\omega_{\phi}\geq0$)
for a certain functional $\mathcal{F}_{\infty}$ on $\mathcal{C}^{\infty}(X),$
whose Legendre transform is $E(\mu).$ The key point, as shown in
\cite{b-b}, is that $\mathcal{F}_{\infty}$ is one time differentiable
on $\mathcal{C}^{\infty}(X),$ which hence establishes the absence
of a phase transition with respect to perturbations of $\phi$ for
the $\beta_{k}=2-$ensemble. 

Incidentally, as explained in \cite{berm3} the large deviation asymptotics
\ref{eq:large deve as beta 2} can, from a physical point of view,
be interpreted as an \emph{effective bosonization }of a fermionic
quantum field theory on $X$ (but is should be pointed out that this
is only an interpretation: no bosonization is actually used in the
derivation \ref{eq:large deve as beta 2} as explained above). In
other words, the collective theory of $N$ fermions is effectively
described by a bosonic field theory, as $N\rightarrow\infty.$ The
point is that the asymptotics \ref{eq:large deve as beta 2} is equivalent,
at least formally, to the asymptotics 

\begin{equation}
\left\langle \left\Vert \Psi(x_{1})\right\Vert ^{2}\cdots\left\Vert \Psi(x_{N})\right\Vert ^{2}\right\rangle \sim\left\langle e^{i\phi(x_{1})}\cdots e^{i\phi(x_{N})}\right\rangle ,\label{eq:boson ans}\end{equation}
in the large $N-$limit. Here the lhs above is the usual $N-$point
function for a fermionic quantum field theory with the usual gauged
Dirac action, i.e. it is the following functional integral over Grassman
fields:

\begin{equation}
\left\langle \left\Vert \Psi(x_{1})\right\Vert ^{2}\cdots\left\Vert \Psi(x_{N})\right\Vert ^{2}\right\rangle :=\int\mathcal{D}\Psi\mathcal{D}\bar{\Psi}e^{-S_{ferm}(\Psi,\bar{\Psi})}\left\Vert \Psi(x_{1})\right\Vert ^{2}\cdots\left\Vert \Psi(x_{1})\right\Vert ^{2},\label{eq:fermion path int}\end{equation}
integrating of over all complex spinors, i.e. smooth sections of the
exterior algebra $\Lambda^{0,*}(T^{^{*}}X)\otimes L^{\otimes k}$
and where $S_{ferm}(\Psi,\bar{\Psi})$ is the fermionic action \[
S_{ferm}(\Psi,\bar{\Psi})=\int_{X}\left\langle \D_{kA}\Psi,\Psi\right\rangle \mu_{0},\]
 expressed in terms of the gauged Dirac operator $\D_{kA}$ on $\Lambda^{0,*}(T^{^{*}}X)\otimes L^{\otimes k}$
induced by the complex structure $J,$ the Hermitian metric on $L,$
i.e. a gauge field $A$ and a choice of Hermitian metric $g$ on $X$
with volume form $\mu_{0},$ i.e. $\D_{kA}=\overline{\partial}+\overline{\partial}^{*}$
a (see \cite{b-v-} for the Riemann surface case). The integer $N$
is the dimension of the space of zero-modes of $\D_{A}$ on $\Lambda^{0,*}(T^{^{*}}X)\otimes L^{\otimes k}$
which coincides with $H^{0}(X,L^{\otimes k})$ due to Kodaira vanishing
in positive degrees (when $k>>1)$ and is hence, in fact, independent
of the metric $g.$ As for the rhs in \ref{eq:boson ans} the bracket
denote integration wrt the (formal) functional measure $\mathcal{D}\phi e^{-S_{bose}(\phi)}$
over all scalar field $\phi$ on $X$ and where \begin{equation}
S_{bose}(\phi)=-\frac{1}{(-i)^{n-1}}\mathcal{E}_{-i\omega}(\phi)\label{eq:ansats for bos act}\end{equation}
defining bosonic action which is a {}``higher-derivative action''
when $n>1$ and when $n=1$ it coincides with non-solotonic part of
the bosonic action obtained in \cite{b-v-,vv}.

\section{Relation to the Yau-Tian-Donaldson program and quantum gravity}

\subsection{\label{sub:Duality-and-relation}Duality and relation to the Yau-Tian-Donaldson
program and balanced metrics }

In this section we will briefly point out some relations to the influential
Yau-Tian-Donaldson program in Kähler geometry \cite{do1,don1,t,p-s}.
In a nutshell the idea of this program is to approximate Kähler-Einstein
metrics (and more general extremal metrics), by a limit of finite
dimensional objects of an algebro-geometrical nature. There are various
versions of this program, but the one which is most relevant for the
present paper is Donaldson's notion of\emph{ canonically balanced
metrics} introduced in \cite{don2}, which is particularly adapted
to Kähler-Einstein metrics (as opposed to general extremal metrics). 

To highlight the similarities let us first formulate a more general
{}``$\beta-$analogue'' of Donaldson's setting, starting with an
ample line bundle $L\rightarrow X.$ The main point is to replace
the infinite dimensional space $\mathcal{H_{\omega}}(X)$ of Kähler
potentials for $\omega\in c_{1}(L)$ (we recall that $\mathcal{H_{\omega}}(X)$
may be identified with the space of all positively curved metrics
on $L)$ with its \emph{quantization at level $k.$ }This latter space,
denoted by $\mathcal{H}_{k},$ is the space of all Hermitian metrics
on the finite dimensional vector space $H^{0}(X,L^{\otimes k}).$
Upon fixing a reference metric\emph{ }$\mathcal{H}_{k}$ is hence
isomorphic to the symmetric space \begin{equation}
GL(N_{k})/U(N_{k})\label{eq:symmetric space finite}\end{equation}
 of all Hermitian $N_{k}\times N_{k}$- matrices. There is a natural
injection defined by the \emph{Fubini-Study map }$FS_{k}(H)$ at level
$k:$ \begin{equation}
FS_{k}:\,\mathcal{H}_{k}\rightarrow\mathcal{H_{\omega}}(X),\,\,\, FS_{k}(H)(x):=\frac{1}{k}\log\sum_{I=1}^{N}\left\Vert \Psi_{I}(x)\right\Vert ^{2},\label{eq:fs}\end{equation}
 expressed in terms of the point-wise norms with respect to the fixed
metric $h_{0}^{\otimes k}$ on $L^{\otimes k}$ of a base $(\Psi_{I})$
in $H^{0}(X;L^{\otimes k})$ which is orthonormal with respect to
$H.$ Moreover, for any given $\beta$ we may define a map in the
reversed direction that we will call $Hilb_{k,\beta}:$ \[
Hilb_{k,\beta}:\,\mathcal{H_{\omega}}(X)\rightarrow\mathcal{H}_{k}\]
 defined as follows: $Hilb_{k,\beta}(\phi)$ is the Hermitian product
(or equivalently, Hilbert norm) on $H^{0}(X,L^{\otimes k})$ defined
by \[
\left\langle f,g\right\rangle _{Hilb_{k,\beta}(\phi)}:=\int_{X}\left\langle f,g\right\rangle e^{-k\phi}e^{\beta\phi}\mu_{0}\]
(note that $\left\langle \cdot,\cdot\right\rangle e^{-k\phi}$ is
the Hermitian metric on $L^{\otimes k}$ \emph{naturally }associated
to $\phi\in\mathcal{H}_{\omega}(X)$ and the remaining factor $e^{\beta\phi}\mu_{0}$
should be thought of as a specific choice of integration element depending
on $\phi).$ An element $H_{k}$ in $\mathcal{H}_{k}$ will be said
to be\emph{ $\beta-$balanced at level $k$} with respect to $(\omega,\mu_{0})$
if is is a fixed point under the composed map \begin{equation}
T_{k,\beta}:=Hilb_{k,\beta}(\phi)\circ FS_{k}:\,\,\,\mathcal{H}_{k}\rightarrow\mathcal{H}_{k}.\label{eq:donaldsons t map}\end{equation}
 on $\mathcal{H}_{k}.$ Equivalently, this means that $H_{k}$ is
a critical point of the following functional $\mathcal{G}_{k}$ on
$\mathcal{H}_{k}:$ \[
\mathcal{G}_{k}(H):=-\frac{1}{kN}\log\det(H)-\frac{1}{\beta}\log\int_{X}e^{\beta FS_{k}(H)}\mu_{0},\]
(after normalization). Repeating the arguments in the proof of Theorem
7.1 in \cite{bbgz} concerning the canonical case when $L=K_{X}$
((the case referred to as $S_{+}$ in \cite{bbgz}) essentially word
for word, one obtains the existence and uniqueness of a $H_{k}\in\mathcal{H}_{k}$
which is \emph{$\beta-$balanced at level $k$} with respect to $(\omega,\mu_{0})$
and such that \[
FS_{k}(H)\rightarrow u_{\beta},\]
in $\mathcal{H_{\omega}}(X)$ when $k\rightarrow\infty$ (or equivalently,
$N\rightarrow\infty)$ where $u_{\beta}$ is the unique solution of
the Monge-Ampère mean field equation \ref{eq:m-a mfe ej z}, assuming
$\beta>0.$ As explained in \cite{bbgz} the main point of the proof
is to show that any limit point in $\mathcal{H_{\omega}}(X)/\R$ of
the sequence $FS_{k}(H)$ is a maximizer of the following functional
on $\mathcal{H_{\omega}}(X):$ \[
\mathcal{G}(\phi):=\mathcal{E}(\phi)-\frac{1}{\beta}\log\int_{X}e^{\beta\phi}\mu_{0},\]
 whose critical points are precisely the solutions of the Monge-Ampère
mean field equation \ref{eq:mean field equation}. Note that the functional
$\mathcal{G}$ is invariant under the natural action by $\R,$ $\phi\rightarrow\phi+c$
and hence maximizing the functional \begin{equation}
\mathcal{E}(\phi)-\frac{1}{\beta}\int_{X}e^{\beta\phi}\mu_{0}\label{eq:liov exp}\end{equation}
picks out the maximizers of $\mathcal{G}$ which satisfies the normalization
\[
\int_{X}e^{\beta\phi}\mu_{0}=1\]
In the Riemann surface case the functional \ref{eq:liov exp} with
the exponential term is also sometimes referred to as the\emph{ Liouville
action} (it appears for example in Polyakov's  functional integral
quantization of the bosonic string, further developed in \cite{o-v}) 

To see the relation to the $\beta-$ensembles introduced in section
\ref{sub:General-ensembles.} and their thermodynamical limit one
should keep in mind the basic linear duality between \emph{functions
}$\phi$ and \emph{measures} $\mu$ defined by the basic pairing \ref{eq:pairing}
In turn, this pairing induces, using the Legendre transform a non-linear
duality between\emph{ convex functionals} of $\phi$ on one hand and
convex functionals of $\mu,$ on the other.

The roles of the spaces $\mathcal{H_{\omega}}(X)$ and $\mathcal{H}_{k}(X)$
are now played by the space $\mathcal{M}_{1}(X)$ and $\mathcal{M}(X^{N_{k}}),$
respectively, where $\mathcal{M}_{N_{k}}(X)$ denotes the space of
all symmetric probability measures on the product $X^{N}$ (i.e. all
$N_{k}-$particle random point processes on $X).$ The analogue of
the Fubini-Study map \ref{eq:fs} is the map \[
\mathcal{M}(X^{N_{k}})\rightarrow\mathcal{M}_{1}(X),\,\,\,\mu_{N}\mapsto(\mu_{N})_{1}:=\left\langle \frac{1}{N}\sum_{i}\delta_{x_{i}}\right\rangle ,\]
 sending a random point process to its one-point correlation measure.
Finally, the role of a $\beta-$balanced metric is now played by the
measure $\mu^{(N_{k})}\in\mathcal{M}(X^{N_{k}})$ defining the $\beta-$ensemble
with $N_{k}$ particles, which was expressed as a Boltzmann-Gibbs
ensemble with Hamiltonian $H^{(N_{k})}$ in formula \ref{eq:bolt-gibbs}.
The point is that $\mu^{(N_{k})}$ can also be defined by a variational
principle. Indeed, by the $N-$particle Gibbs principle for canonical
ensembles $\mu_{(N_{k})}$ is the unique minimizer of the $N-$particle
mean free energy functional on $\mathcal{M}(X^{N_{k}}):$ \begin{equation}
F^{(N)}(\mu_{N})=\frac{1}{N}\int_{X^{N}}\mu_{N}H^{(N)}-\frac{1}{N}S(\mu_{N},\mu_{0}^{\otimes N})\label{eq:N particle free en}\end{equation}
i.e. the difference between\emph{ mean energy} and \emph{mean entropy}.
There is also an analogue of the definition of a balanced metric as
a fixed point of the map $T_{k,\beta}$ above. Indeed, it is well-known
that any Gibbs-Boltzmann measure can be uniquely determined as a stationary
state for a stochastic process $\mu_{t}$ on $X^{N}$ defined by suitable
Glauber (or Langevin) dynamics, but we will not develop this point
of view here. 

Interestingly, performing a Legendre transform of each of the two
convex functionals on $\mathcal{M}_{1}(X)$ summing up to the free
energy functional $F(\mu)$ (i.e. the pluricomplex energy $E(\mu)$
and minus entropy $-\frac{1}{\beta}S(\mu))$ yields a functional on
$\mathcal{H_{\omega}}(X)$ which is nothing but the functional $\mathcal{G}$
above: \[
F=E+(-\frac{1}{\beta}S),\,\,\,\,\mathcal{G}=E^{*}+(-\frac{1}{\beta}S)^{*}\]
It should be pointed out that in the canonical case (where the critical
points of the functionals are Kähler-Einstein metrics) the two functionals
$F$ and $\mathcal{G}$ have already appeared in Kähler geometry from
a different point of view. For example, the limiting free energy functional
$F(\mu)$ on $\mathcal{M}_{1}(X)$ may be identified with Mabuchi's
$K-$energy $\nu$ of a Kähler metric in $c_{1}(K_{X}):$ \begin{equation}
F(\omega_{\phi}^{n}/Vn!)=\nu(\omega_{\phi})\label{eq:f as mab}\end{equation}
The functional $\nu$ was first introduced by Mabuchi as the functional
on $\mathcal{H_{\omega}}(X)$ whose gradient with respect to the Mabuchi-Semmes-Donaldson
Riemannian metric on $\mathcal{H_{\omega}}(X)$ is the scalar curvature
minus its average \cite{t,p-s}. But this is easily seen to be equivalent
to the variational property \ref{eq:variational der of f} of $F$
and hence $F$ and $\nu$ coincide under the identification above
(up to an additive constant). The explicit formula for $\nu$ obtained
from the identification \ref{eq:f as mab} is in fact equivalent to
an explicit formula for $\nu$ of Tian and Chen \cite{t}. Moreover,
the functional $-G$ coincides with the Ding called functional (see
the book \cite{t} and references therein). Interestingly, using Legendre
transforms as above one arrives at new proofs and generalizations
of various useful results in Kähler geometry \cite{berm6}. 

Finally, it seems worth pointing out that in the case when $\beta=0$
the notion of balanced metrics still makes sense and was studied by
Donaldson in \cite{don2} with a particular emphasize on the case
when $X$ is a Calabi-Yau form. Then $\mu$ may be canonically chosen
as $i^{n^{2}}\Omega\wedge\bar{\Omega}/\int i^{n^{2}}\Omega\wedge\bar{\Omega}$
where $\Omega$ is non-vanishing holomorphic $n-$form on $X$ and
the curvature forms of the balanced metrics at level $k$ then converge
to the unique Ricci flat metric in $[\omega],$ whose existence was
established in Yau's proof of the Calabi conjecture. Relation between
these balanced metrics on Calabi-Yau manifolds and black holes were
considered in \cite{d-k2}. However, in the case when $\beta=0$ the
$\beta-$ensembles introduced in the present work appear to be less
interesting: they are pure Poisson processes without any connections
to fermions. The considerably more complicated case when $\beta$
is \emph{negative} is briefly discussed below in connection to Kähler-Einstein
metrics with positive Ricci curvature.

\subsection{Relations to quantum gravity and the space of all complex structures}

In Hawking's approach to Euclidean quantum gravity \cite{ha} one
considers the (formal) functional measure \[
\mathcal{D}ge^{-S(g)}\]
on the space of all Riemannian metrics $g$ (modulo diffeomorphisms),
where $S(g)$ is the Einstein-Hilbert action with a cosmological constant
$\Lambda$ and where we have set the fundamental constants to be equal
to one (but eventually we will discuss semi-classical limits). One
also has to integrate over all diffeomorphism types of four-manifolds,
but the relation to the present setting appears more closely when
one considers the restriction of the integration to metrics on a fixed
compact smooth manifold $X.$ As discussed by Hawking, by choosing
a conformal gauge, the integration may be decomposed as an integration
over a conformal factor and the space of conformal equivalence classes
of metrics (see section $5$ in \cite{ha} and also the very recent
paper \cite{hoo} of t' Hooft ).

To see the relation to the present setting we let $X$ be a smooth
oriented manifold admitting some complex structure $J_{0}$ such that
the corresponding canonical line bundle $K_{X_{0}}$ is ample (when
$n=1$ this just means that $X$ has genus at least two). For simplicity
we will also assume that $H^{1}(X,\R)=0.$ We let $\mathcal{J}(X)$
be the space of \emph{all} complex structures $J$ on $X,$ which
are compatible with the orientation. Given $J$ we denote by $\mathcal{K}(X,J)$
be the space of all Kähler metrics on $(X,J)$ in the cohomology class
$c_{1}(K_{(X,J)})$ and we let $\mathcal{K}(X)$ be their union over
all $J$ in $\mathcal{J}(X).$ The group $DIFF(X)$ of diffeomorphisms
on $X$ acts naturally on $\mathcal{J}(X)$ and $K(X)$ and we let
$\mathcal{T}(X)$ be the corresponding moduli space $\mathcal{T}(X)=\mathcal{J}(X)/DIFF(X)$
\footnote{to avoid technical difficulties one should really replace $DIFF(X)$
with its subgroup $DIFF_{0}(X)$ in the connected component of the
identity and accordingly the corresponding moduli space with the\emph{
Teichmuller space} associated to $X$ but to simplify the discussion
we will not make this distinction. For mathematical results about
the relevant moduli spaces we refer the reader to \cite{fs,sc}. %
}and consider the associated bundle \begin{equation}
\mathcal{K}(X)/DIFF(X)\rightarrow\mathcal{T}(X)\label{eq:bundle of equiv classes of kahler me}\end{equation}
of equivalence classes of Kähler metrics. This bundle may be given
a useful complex geometric realization as follows. First we recall
that there is a natural holomorphic fibration (the \emph{universal
family}) \[
\mathcal{X}\rightarrow\mathcal{T}(X)\]
such the fiber $\mathcal{X}_{[J]}$ is a complex manifold biholomorphic
to $(X,J)$ (when $n=1$ $\mathcal{X}$ is called the \emph{universal
curve} over $\mathcal{T}(X)).$ The bundle \ref{eq:bundle of equiv classes of kahler me}
may now be identified with an infinite dimensional bundle of metrics
over $\mathcal{T}(X)$ such that the fiber over $[J]$ gets identified
with the corresponding space of Kähler metrics on $\mathcal{X}_{[J]}.$
Equivalently, \[
\mathcal{K}(X)/DIFF(X)\simeq\mathcal{H}(K_{\mathcal{X}/\mathcal{T}(X)})/\R,\]
where the rhs is the space of all fiber-wise positively curved metrics
on the relative canonical line bundle $K_{\mathcal{X}/\mathcal{T}(X)}$
over $\mathcal{X}:$ \[
K_{\mathcal{X}/\mathcal{T}(X)}\rightarrow\mathcal{X}\rightarrow\mathcal{T}(X),\]
 (by definition the restriction of $K_{\mathcal{X}/\mathcal{T}(X)}$
to each fiber is the usual canonical line bundle). We can now consider
a (formal) functional measure on the space $\mathcal{K}(X)/DIFF(X)$
of the form \begin{equation}
\mathcal{D}ge^{-\frac{1}{N}F(g)}\label{eq:functional measure on space of kahler}\end{equation}
where the positive integer $N$ plays the role of a semi-classical
parameter and we simply let \[
F(g):=F(\mu_{g}):=F(\frac{dVol_{g}}{V})\]
where $F(\mu)$ is the free energy functional defined in section \ref{sub:Convergence-in-the thermo}
and where we have used that $g$ may be identified with a Kähler metric
$\omega_{g}$ in $c_{1}(K_{\mathcal{X}_{[J]}})$ for some $J,$ so
that $dVol_{g}=\omega_{g}^{n}/n!.$ In other words, fiber-wise over
$\mathcal{T}(X)$ the functional $F(g)$ may be identified with the
free energy functional for a fixed complex manifold dimorphic to $X$
(which in turn as explained in the previous section may be identified
with Mabuchi's $K-$energy). Using the previous arguments we can now
give a well-defined and canonical {}``finite $N$ approximation''
to the functional measure \ref{eq:functional measure on space of kahler}
as follows. First note that for each fiber $\mathcal{X}_{[J]}$ we
have the canonical probability measure $\mu^{(N)}$ on $\mathcal{X}_{[J]}^{N}$
(defined in terms of fermions in section \ref{sub:The-canonical-background}).
Moreover the base $\mathcal{T}(X)$ admits a canonical metric, the
Weil-Peterson metric, with finite volume (as follows form the results
in \cite{sc}). Altogether this induces a {}``universal'' probability
measure $\tilde{\mu}^{(N)}$ on the finite-dimensional complex manifold
\[
\mathcal{X}^{(N)}:=\bigcup_{[J]\in\mathcal{T}(X)}\mathcal{X}_{[J]}^{N}\]
and the arguments in section \ref{sub:Convergence-in-the thermo}
give the asymptotics \[
\mathcal{D}ge^{-\frac{1}{N}F(g)}\sim\tilde{\mu}^{(N)}\]
 as $N\rightarrow\infty$ in a suitable sense, so that the measure
$\tilde{\mu}^{(N)}$ is exponentially concentrated on the moduli space
of all Einstein metrics on $X$ which are Kähler for some complex
structure. To see the relation to Hawking's setting one may think
of $\mathcal{T}(X)$ as playing the role of the space of all conformal
equivalence classes and the Kähler potentials as the logarithm of
the conformal factors. This is more then an analogy when $n=1$ since
the space $\mathcal{K}(X)$ then coincides with the space of all Riemannian
metrics on $X,$ normalized so that their total area is $2\mbox{genus}(X)-2.$
However, in higher dimensions non-trivial integrability conditions
appear. Of course, another difference is that the action $F(g)$ is
not coming form the Einstein-Hilbert action (in any dimension). As
explained in the previous section the action is, in a sense, dual
to an action generalizing the Liouville action to higher dimension.

Finally, following Fujiki \cite{fu} and Donaldson \cite{do3} we
will explain how the action $F$ appears naturally from a symplecto-geometric
point of view when the complex structures $J$ are used as dynamical
variable instead of Riemannian metrics (we refer to \cite{fu,do3}
for more details and further references). Fix a symplectic form $\omega$
in the real cohomology class $-c_{1}(X)=$: $c_{1}(K_{(X,J_{0})}))$
in $H^{2}(X,\Z)$ for any $J\in\mathcal{J}(X)$ and consider the subspace
$\mathcal{J}(X,\omega)\subset\mathcal{J}(X)$ of all complex structures
$J$ on $X$ which are compatible with $\omega,$ i.e. such that \begin{equation}
\mathcal{J}(X,\omega)\rightarrow\mathcal{G}(X,\omega):\,\,\, J\mapsto g_{(\omega,J)}:=\omega(\cdot,J\cdot)\label{eq:map j to g}\end{equation}
 maps $J$ to a Riemannian metric which is Kähler wrt $J$ and with
$\omega$ as its Kähler form. The map \ref{eq:map j to g} naturally
defines a principal $SDIFF(X,\omega)-$bundle over $\mathcal{G}(X,\omega),$
where $SDIFF(X,\omega)$ is the symplectomorphism group, i.e. $SDIFF(X,\omega)$
acts freely by pull-back on the fibers so that \[
\mathcal{J}(X,\omega)/SDIFF(X,\omega)\simeq\mathcal{G}(X,\omega)\]
 The infinite dimensional space $\mathcal{J}(X,\omega)$ is itself
naturally a Kähler manifold with a Weil-Petersson type Kähler form
$\Omega$ such that $SDIFF(X,\omega)$ acts by holomorphic symplectomorphism
on $(\mathcal{J}(X,\omega),\Omega).$ More precisely, the action lifts
to an Hermitian line bundle over $\mathcal{J}(X,\omega)$ with curvature
$\Omega$ can be realized as a certain determinant line bundle (or
alternatively as a Deligne pairing \cite{p-s}). This allows one to
apply the powerful formalism of {}``moment maps'' for group actions
on Kähler manifolds and geometric invariant theory (in this case the
corresponding moment map turns out to be represented by the scalar
curvature). Even though the complexified group $SDIFF(X,\omega)_{\C}$
does not exist, in the strict sense, it is still possible to give
a meaning to its orbits in $\mathcal{J}(X,\omega)$ which are complex
submanifolds of $\mathcal{J}(X,\omega).$ On each orbit the Kähler
metric $\Omega$ has an $SDIFF(X,\omega)$- invariant Kähler potential
$F(J)$ which is defined up to an additive constant (corresponding
to the {}``norm-functional''/ {}``Kempf-Ness functional'' in the
abstract setting of moment maps). The functional $F(J)$ may be uniquely
determined by requiring its infimum to be equal to $0$ (in general
the {}``norm functional'' is bounded from below precisely when the
group action is semi-stable which is indeed the case in our setting).
Finally, for any fixed complex structure $J_{0}$ the quotient of
the orbit of $SDIFF(X,\omega)_{\C}$ passing through $J_{0},$ i.e.
\begin{equation}
\left(SDIFF(X,\omega)_{\C}J_{0}\right)/SDIFF(X,\omega)\label{eq:infinite dim symm space}\end{equation}
 may, under the map \ref{eq:map j to g}, be identified with the space
of all Kähler metrics in $c_{1}(K_{(X,J_{0})})$ (this is the $N=\infty$
analogue of \ref{eq:symmetric space finite}). Under this identification
$F(J)$ coincides with the corresponding free energy functional (since,
as shown by Fujiki and Donaldson it coincides with Mabuchi's $K-$energy
functional on the space of Kähler metrics). Concretely, the identification
used above is obtained by noting that $J$ is in the orbit $SDIFF(X,\omega)_{\C}J_{0}$
iff there is $f\in DIFF(X)$ such that the pull-back $f^{*}\omega$
defines a Kähler form on $(X,J_{0})$ which is cohomologous to $\omega.$ 

In conclusion this means that from one point of view the space $K(X)/DIFF(X)$
where the functional measure \ref{eq:functional measure on space of kahler}
lives may be identified with the space of positively curved metrics
on the relative canonical line bundle $K_{\mathcal{X}/\mathcal{T}(X)}$
(mod $\R)$ and the action $F(g)$ with the free energy functional,
fiber-wise over the moduli space $\mathcal{T}(X).$ From another point
of view $K(X)/DIFF(X)$ may be identified with $\mathcal{J}(X,\omega)/SDIFF(X,\omega)$
and the corresponding action $F(g)$ with a (normalized) $SDIFF(X,\omega)-$invariant
Kähler potential $F(J)$ on the infinite dimensional Kähler manifold
$\mathcal{J}(X,\omega).$ In other words, when formulating the theory
directly on the space $\mathcal{J}(X,\omega)$ it becomes a gauge
theory with structure group $SDIFF(X,\omega)$ for a bundle over the
moduli space of complex structures. The finite $N-$approximation
is obtained by replacing $K(X)/DIFF(X)$ with the finite dimensional
complex manifold $\mathcal{X}^{(N)}$ fibered over $\mathcal{T}(X)$
and replacing the formal functional measure with a probability measure
on $\mathcal{X}^{(N)}.$ Finally we note that another somewhat dual
approximation, in the spirit of the Yau-Tian- Donaldson program discussed
in the previous section, could also be obtained by instead replacing
$\mathcal{K}(X)/DIFF(X)$ with the space of Hermitian metrics on the
vector bundle $E_{N}\rightarrow\mathcal{T}(X)$ defined as the the
direct image of $K_{\mathcal{X}/\mathcal{T}(X)}^{\otimes k},$ i.e.
\[
E_{N_{k}}:=\bigcup_{[J]\in\mathcal{T}(X)}H^{0}(\mathcal{X}_{[J]},K_{\mathcal{X}_{[J]}}^{\otimes k})\]
(compare the setup in \cite{berm3-1}). Note that in this latter case
the structure group is $U(N)$ and, in a sense, the Yau-Tian- Donaldson
program concerns the problem of making the heuristic statement $U(N)\rightarrow SDIFF(X,\omega),$
as $N\rightarrow\infty,$ mathematically precise.

\subsection{\label{sub:Conclusion-and-discussion}Conclusion and outlook}

To a given a compact manifold with a fixed integrable complex structure
$J$ we have associated a canonical $N-$particle free fermion gas
whose one-particle correlation measures converge in the thermodynamical
(large $N)$ limit to the volume form of the Kähler-Einstein metric
$\omega_{KE}$ associated to $(X,J).$ More precisely, it was assumed
that the canonical line bundle $K_{X}$ be positive (i.e. ample),
which corresponds to $\omega_{KE}$ having \emph{negative} Ricci curvature
and the one-particle quantum state space of the fermion gas was taken
as the $N_{k}-$dimensional space $H^{0}(X,K_{X}^{\otimes k})$ of
global holomorphic sections on $X$ with values in $K_{X}^{\otimes k}.$
The argument in fact gave precise exponentially small fluctuations
around the Kähler-Einstein volume forms with a rate function $F$
naturally identified with Mabuchi's $K-$energy (which plays an important
rule in Kähler geometry). The convergence in the thermodynamical limit
was obtained by introducing an\emph{ auxiliary }background Kähler
form $\omega$ in the first Chern class $c_{1}(K_{X})$ (or equivalently
a metric on $K_{X}$) also determining a volume form on $X.$ This
lead to a decomposition of the rate functional $F$ as \[
F(\mu)=E(\mu)-S(\mu),\]
 (with both terms depending on the choice of $\omega).$ In terms
of the statistical mechanics of a classical canonical Boltzmann-Gibbs
ensemble $E$ and $S$ appeared as the limiting\emph{ mean energy}
and \emph{mean entropy,} respectively. Minimizing $F$ then gave an
equation of mean field type whose unique solution is given by the
Kähler-Einstein volume form. An interpretation of the energy $E$
in terms of \emph{effective bosonization} was also given. 

Heuristically, it seems that one could interpret the thermodynamical
limit above as saying that the Kähler-Einstein metric emerges from
a \emph{fermionic maximum entropy principle}: the particles try to
maximize their entropy (i.e. volume in configuration space) under
the constraint that they behave as fermions and hence, according to
Pauli's exclusion principle, cannot occupy the same space leading
to an effective repulsion. 

By making the complex structure $J$ dynamical relations to a variant
of quantum gravity were explained using a canonical finite $N$ approximation
of the functional measure. It would be interesting to compare with
other discretization schemes such as spin foam loop quantum gravity
or the approach of CDT (causal dynamical triangulations); see \cite{n-p,lo,a---}
and references therein. However, in the present case one only consider
fluctuating Riemannian metrics which are Kähler with respect to some
complex structure on $X$ (imposing a constraint unless $n=1).$ 

The precise mathematical details of the thermodynamical convergence
will be investigated elsewhere, as well as the case when the anti-canonica
line bundle $K_{X}^{-1}$ is positive (so that any Kähler-Einstein
metric must have positive Ricci curvature). In the latter case there
are well-known obstructions to the existence of Kähler-Einstein metrics
and the Yau-Tian-Donaldson program aims at showing that all obstruction
may be formulated in terms of algebro-geometric stability. More precisely,
according to the the Yau-Tian-Donaldson conjecture the existence of
Kähler-Einstein is equivalent $X$ beeing $K-$stable (or to some
refined notion of $K-$stability) \cite{p-s}. From the point of view
of the present paper this is related to the fact that the natural
candidate for the $N-$particle ensemble in the case when $K_{X}^{-1}$
is positive may be formally written as \begin{equation}
\mu^{(N)}=(\Psi_{1}(x_{1})\wedge\bar{\Psi}_{1}(x_{1})\wedge\cdots\wedge\Psi_{N}(x_{N})\wedge\bar{\Psi}_{N}(x_{N}))^{-1/k}/\mathcal{Z}_{N}\label{eq:prob measure for kx neg}\end{equation}
 where now $(\Psi_{I})$ is a base in $H^{0}(X,(K_{X}^{-1})^{\otimes k}).$
However, because of the negative exponent above this singular volume
form on $X^{N}$ may be non-integrable, i.e. $\mathcal{Z}_{N}=\infty,$
due to singularities appearing when to points merge. This means that
the correponding Boltszmann-Gibbs measure $\mu^{(N_{k})}$ at level
$k$ may not even exist. In fact, its existence should be closely
relatated to the existence of balanced metrics, which in turn is closely
related to the notion of $K-$stability (compare the disucssion in
section \ref{sub:Duality-and-relation}). Anyway, fixing an auxiliary
smooth Hermitian metric $h=e^{-\Phi_{0}}$ on $K_{X}^{-1}$ one can
look at the $\beta-$ensemble \begin{equation}
\mu_{\beta}^{(N)}=\left|\Psi_{1}\wedge\bar{\Psi}_{1}\wedge\cdots\wedge\Psi_{N}\right|^{\beta/k}((e^{\Phi})^{\otimes N}){}^{(-2-\beta)}/\mathcal{Z}_{\beta,N}\label{eq:renormalized prob meas}\end{equation}
 (with $\beta=-\gamma$ \emph{negative}) which is integrable for $\gamma$
sufficiently small, coinciding with \ref{eq:prob measure for kx neg}
for $\beta=2.$ In fact, it can be shown that the measure is integrable
as long as $\gamma<\alpha_{X},$ where $\alpha_{X}$ is Tian's\emph{
$\alpha-$invariant} \cite{t} (also called the log canonical threshold
in algebraic geometry ). Repeating the same argument as before the
rate functional in \ref{eq:rasympt with rate fun} may be written
as \begin{equation}
\beta F_{\beta}(\mu)=\beta E(\mu)-S(\mu)\label{eq:free energy times beta in neg}\end{equation}
In the case of main interest, i.e. \textbf{$\beta=-2,$} this functional
may be identified with Mabuchi's $K-$energy functional in Kähler
geometry. Since now $\beta<0$ this means that the energy contribution
$\beta E(\mu)$ corresponds to an \emph{attractive} force. Hence,
minimizimgs $\beta F_{\beta}(\mu)$ amounts to finding a balance between
\emph{minimizing energy} (which brings the particles close together)
and \emph{maximizing entropy} (which rather has a repelling effect).
Accordingly, if $\beta$ is too small then the attractive effect wins
and the rate funcional \ref{eq:free energy times beta in neg} becomes
unbounded from below. Interestingly, it can be shown that when $X$
admits a Kähler-Einstein metrics with volume form $\mu_{KE}$ and
holomorphic vector fields (so that the metric is non-unique) then
the rate functional \ref{eq:free energy times beta in neg} is bounded
from below for $\beta\geq-2$ (with minimizer $\mu_{KE}$ for $\beta=-2)$
and it is unbounded from below when $\beta<-2.$ In fact, the notion
of\emph{ analytic} $K-$stability which by Tian's theorem is equivalent
to the existence of a unique Kähler-Einstein metric \cite{t} may
be equivalently formulated as the condition that $\beta F_{\beta}$
be bounded from below for all $\beta$ sufficently close to $-2$
(see \cite{berm6} and references therein). As suggested above it
seems natural to expect that the ordinary notion of (non-analytic)
$K-$stability is closely related to integrability of the measure
\ref{eq:prob measure for kx neg} for $N$ sufficently large. In turn,
by Gibbs principle this latter property is equivalent to the boundedness
from below of the functional $\beta F^{(N)},$ where $F^{(N)}$ is
the corresponding $N-$particle Gibbs free energy (formula \ref{eq:N particle free en}).

If $(X,J)$ admits a Kähler-Einstein metric (or if appropriate stability
conditions hold) it seems natural to expect the one-point correlation
measure of $\mu_{\beta}^{(N)}$ to converge to a Kähler-Einstein volume
form when first $N\rightarrow\infty$ and then $\beta\rightarrow-2.$
The simplest case appears when $X$ is the Riemann sphere. Then the
corresponding ensemble is explicitly given by a one component plasma
(or equivalently a point vortex system) studied by Kiessling in \cite{ki2}
where a first order phase transition appears at $\beta=-2.$ Kiessling
also considered higherdimensional generalizations in other directions
than the one explored here, namely to the conformal geometry of higher
dimensional spheres where the correponding Hamiltonian $H^{(N)}$
is a sum of logarithmic pair interaction. As a consequence the corresponding
mean field equations are quasi-linear (with the non-linearity coming
from the exponential term), as opposed to the present setting where
the fully non-linear Monge-Ampère operator appears.

There is also an interesting variant of the probability measure $\mu_{\beta}^{(N)}$
above where one, in the definition \ref{eq:renormalized prob meas},
instead uses a \emph{singular }metric on $K_{X}^{-1}$ whose curvature
is the current $\delta_{D}$ defined by a given Calabi-Yau submanifold
$D$ of dimenion $n-1$ in $X$ (such $D$ usually exist); more precisely,
$D$ is cut out by a holomorphic section $s$ of $K_{X}^{-1}$ and
one takes $\Phi=\log|s|^{2}$ in \ref{eq:renormalized prob meas}.
Then the minimizers $\mu_{\gamma}(=\omega_{\gamma}^{n}/n)$ of the
corresponding scaled free energy functional $\beta F_{\beta,D}(\mu)$
(now writing $\beta=-2\gamma$ for $0<\gamma<1)$ turn out to satisfy
an equation very recently introduced by Donaldson \cite{do-2} (see
also \cite{do-3}): \begin{equation}
\mbox{Ric}\omega_{\gamma}=\gamma\omega_{\gamma}+(1-\gamma)\delta_{D}\label{eq:donaldsons equ}\end{equation}
saying that $\omega_{\gamma}$ is a Kähler-Einstein metric on $X-D$
whose Ricci curvature is singular along the submanifold (divisor)
$D.$ The equationa above appeared in a program proposed by Donaldson
to attack the Yau-Tian-Donaldson conjecture where the first step consists
in proving the existence of solutions to the previous equation for\textbf{~}$\gamma<<1$
and the last step amounts to, assumping $K-$stabilility (or rather
a variant like $\bar{K}-$stability), use certain balanced metrics
to take the limit $\gamma\rightarrow1$ (corresponding to $\beta\rightarrow-2$
in the present notation) to obtain a bona fida Kähler-Einstein metric
on $X$.%
\footnote{The existence of the metric in \ref{eq:donaldsons equ} $\omega_{\gamma}$
for $\gamma<<1$ has now been proved in \cite{berm6} using a variational
calculus inspired by (but independent of) the statistical mechanical
model proposed in the present paper. Moreover, this result combined
with the very recent paper \cite{jmr} shows that $\omega_{\gamma}$
in fact has\emph{ conical} singularies with angle $2\pi\gamma$ transversally
to $D$ and is smooth in the directions of $D$ as also conjectured
by Donaldson. %
}

In the light of Donaldson's program it seems natural to formulate
the following conjecture: if $X$ is $K-$stable (or say $\bar{K}-$stable
in Donaldson's sense) and $X$ contains a submanifold $D$ as above,
then the corresponding probability measures $\mu_{\beta,D}^{(N)}$
are well-defined for all $\beta<-2$ and the corresponding one-point
correlation measures converge to a Kähler-Einstein volume form when
first $N\rightarrow\infty$ and then $\beta\rightarrow-2.$

Finally, it would also be interesting to detail the Glauber (Langevin)
stochastic dynamics alluded to section \ref{sub:Duality-and-relation}
and investigate a suitable {}``hydrodynamical'' scaling limit (see
for example the recent paper \cite{d-o-r} for relations between Langevin
dynamics and field theories with holomorphic factorization and supersymmetry).
This should lead to a deterministic heat-equation type flow on the
space of all (smooth) probability measures $\mathcal{M}_{1}(X),$
converging towards the Kähler-Einstein volume form. In fact, a {}``dual''
(in the sense of the previous section) scaling limit of Donaldson's
iteration of the map $T_{k,\beta}$ \ref{eq:donaldsons t map} was
shown to converge to the Kähler-Ricci flow in \cite{berm3-1}.

\end{document}